\documentclass[cits]{PoS}
\usepackage{amssymb,latexsym,amsmath}
\newcommand{\eq}[1]{\begin{align} #1 \end{align}}

\title{Bose-Einstein Condensation of Pions}

\ShortTitle{Bose-Einstein Condensation of Pions}

\author{\speaker{Viktor Begun}
 \\
 Bogolyubov Institute for Theoretical Physics, Kiev, Ukraine.
        \\ E-mail: \email{viktor.begun@gmail.com}}

\author{Mark Gorenstein\\
\\
 Bogolyubov Institute for Theoretical Physics, Kiev, Ukraine.
        \\ E-mail: \email{mark@mgor.kiev.ua}}

\abstract{Particle number fluctuations are studied in the ideal
pion gas approaching Bose-Einstein condensation. Two different
cases are considered: Bose condensation of pions at large charge
densities $\rho_Q$ and Bose condensation at large total densities
of pions $\rho_{\pi}$. Calculations are done in grand canonical,
canonical and microcanonical ensembles. At high collision energy,
in the samples of events with a fixed number of all pions,
$N_{\pi}$, one may observe a prominent signal. When $N_{\pi}$
increases the scaled variances for particle number fluctuations of
both neutral and charged pions increase dramatically in the
vicinity of the Bose-Einstein condensation line. As an example,
the estimates are presented for $p+p$ collisions at the beam
energy of 70~GeV. }

\FullConference{Critical Point and Onset of Deconfinement - 4th International Workshop\\
         July 9 - 13, 2007\\ Darmstadt, Germany}

\begin{document}

Long time ago in 1924 the Bose statistics was discovered
\cite{Bose}, and one year later the phenomenon of Bose-Einstein
condensation (BEC) \cite{Einstein} was predicted. Tremendous
efforts were required however to confirm BEC experimentally. The
atomic gases are transformed into a liquid or solid before
reaching the BEC point. The only way to avoid this is to consider
extremely low densities. At these conditions the thermal
equilibrium in the atomic gas is reached much faster than the
chemical equilibrium. The life time of the 
metastable gas phase is stretched to seconds or minutes. This is
enough to observe the BEC signatures. Small density leads,
however, to small temperature of BEC. Only in 1995 two
experimental groups succeeded to create the `genuine' BE
condensate  by using new developments in cooling and trapping
techniques \cite{cold_atoms}. Leaders of these two groups,
Cornell, Wieman, and Ketterle, won the 2001 Nobel Prize for this
achievement.

Pions are spin-zero mesons. They are the lightest hadrons
copiously produced in high energy collisions. In the present
letter we argue that the pion number fluctuations may give a
prominent signal of approaching the BEC point. In fact, there is
the BEC line in a plane of pion density and temperature. The pion
system should be in a state of thermal, but not chemical,
equilibrium to reach the BEC line. This can be achieved by
selecting the samples of events with high pion multiplicities.
Multipion states are formed in high energy nucleus-nucleus
collisions, as well as in the elementary particle ones. There were
several suggestions to search for BEC of $\pi$-mesons (see, e.g.,
Ref.~\cite{pion-BC}). However, complete statistical mechanics
calculations of pion number fluctuations have never been
presented. There is a qualitative difference in properties of the
mean multiplicity and of the scaled variance of multiplicity
fluctuations in different statistical ensembles. The results
obtained with grand canonical ensemble (GCE), canonical ensemble
(CE), and microcanonical ensemble (MCE) for the mean multiplicity
approach to each other in the large volume limit. This reflects
the thermodynamic equivalence of the statistical ensembles.
Recently it has been found  \cite{CE,BF,PBEC,MCE,CLT} that
corresponding results for the scaled variance are different in
different ensembles, and this difference is preserved in the
thermodynamic limit. To extract the matter properties from
analysis of event-by-event fluctuations, one needs to fix the
samples of high energy events, and choose the corresponding
statistical ensemble for their analysis. This is discussed below
(see also \cite{PBEC}).


Let us start with a well known example of non-relativistic ideal
Bose gas. The occupation numbers, $n_{\bf{p}}$, of single quantum
states, labelled by 3-momenta $\bf{p}$, are equal to
$n_{\bf{p}}=0,1,\ldots,\infty$. In the GCE their average values,
fluctuations, and correlations are the following  \cite{lan}:
\eq{
 \langle n_{\bf{p}} \rangle
  = \frac {1} {\exp \left[\left(\frac{{\bf p}^2}{2m} - \mu \right)/ T\right]
 - 1},
 &&\langle\left(\Delta n_{\bf{p}}\right)^2\rangle
 = \langle n_{\bf{p}}\rangle \left(1
+ \langle n_{\bf{p}} \rangle\right)\equiv \upsilon^{2}_{\bf{p}},
 && \langle \Delta n_{\bf{p}}  \Delta n_{\bf{k}} \rangle
 =\upsilon_{\bf{p}}^2~\delta_{\bf{p}\bf{k}}, %
\label{mcc-gce}
} where $\Delta n_{\bf{p}}\equiv n_{\bf{p}}-\langle
n_{\bf{p}}\rangle$, $m$ denotes the particle mass, $T$ and $\mu$
are the system temperature and chemical potential, respectively
(throughout the paper we use the units with $\hbar=c=k=1$).
The average number of particles in the GCE reads \cite{lan}:
\eq{
\langle N\rangle~ &\equiv~ \overline{N}(V,T,\mu)~=~\sum_{\bf{p}}
\langle n_{\bf{p}} \rangle
 ~=~\frac{V}{2\pi^2}\int_0^{\infty}\frac{p^2dp}
 {\exp\left[\left(\frac{p^2}{2m}~-~\mu \right)/T\right]~-~1}~,
\label{N-nonrel}
}
where $V$ is the system volume. We consider particles with spin
equal to zero, thus the degeneracy factor equals 1.  In the
thermodynamic limit, $V\rightarrow \infty$, the sum over momentum
states is transformed into the momentum integral,
$\sum_{\bf{p}}\ldots =(V/2\pi^2)\int_0^{\infty}\ldots p^2dp$. This
substitution, assumed in all formulae below,  is valid  if the
chemical potential in the non-relativistic Bose gas is restricted
to $\mu < 0$ (or $\mu <m$ in relativistic formulation). When the
temperature $T$ decreases at fixed particle number density $\rho
\equiv \overline{N}/V$, the chemical potential $\mu$ increases and
becomes equal to zero at $T=T_{C}$, known as the BEC temperature.
At this point from Eq.~(\ref{N-nonrel}) one finds,
$\overline{N}(V,T=T_{C},\mu=0)=V[mT_{C}/(2\pi)]^{3/2}\zeta(3/2)$,
where $\zeta(3/2)\cong 2.612$ is the  Riemann zeta-function. This
gives,
\eq{\label{TC-nonrel}
 T_{C}~=~2\pi[\zeta(3/2)]^{-2/3}~\frac{\rho^{2/3}}{m}~\cong~
3.31 ~\frac{\rho^{2/3}}{m}~.
}
At $\mu=0$ and $T<T_C$,  a macroscopic part, $N_{C}$ (called the
BE condensate), of the total particle number occupies the lowest
energy level ${\bf p}=0$. At $\mu=0$ and $T<T_C$ the GCE average
number of particles in the BE condensate is equal to
$N_C=\overline{N}[1-(T/T_C)^{3/2}]$.

Introducing $\Delta N \equiv N-\langle N\rangle$ one finds the
particle number fluctuations in the GCE,
\eq{ 
& \langle (\Delta N)^2 ~\rangle ~=~\sum_{\bf{p},\bf{k}}~\langle
\Delta n_{\bf{p}}~\Delta n_{\bf{k}}\rangle~=~\sum_{\bf{p}}~
v_{\bf{p}}^2~, \quad & \omega \equiv \frac{\langle (\Delta N)^2\,
\rangle}{\langle N \,\rangle} \,=\, \frac{\sum_{\bf{p}}
\upsilon_{\bf{p}}^2}{\sum_{\bf{p}}\langle n_{\bf{p}}\rangle}
~=~1~+~\frac{\sum_{\bf{p}}\langle
n_{\bf{p}}\rangle^2}{\sum_{\bf{p}}\langle n_{\bf{p}}\rangle}~.
\label{omega-nonrel}
}
The limit $-\mu/T\gg 1$ gives $\langle n_{\bf{p}}\rangle\ll 1$.
This corresponds to the Boltzmann approximation, and then from
Eqs.~(\ref{N-nonrel},\ref{omega-nonrel}) it follows:
$\overline{N}(V,T,\mu)\cong V\exp(\mu/T)(mT/2\pi)^{3/2}$ and
$\omega\cong 1$. When $\mu$ increases the scaled variance $\omega$
becomes larger, $\omega>1$. This is the well known Bose
enhancement effect for the particle number fluctuations. From
Eq.~(\ref{omega-nonrel}) at $\mu\rightarrow 0$ one finds
$\omega\rightarrow\infty$. Thus, the anomalous particle number
fluctuations appear in the GCE formulation when the system
approaches the BEC point. Two comments are appropriate here.
First, for finite systems $\omega$ remains finite, and $\omega =
\infty$ emerges from Eq.~(\ref{omega-nonrel}) at $\mu=0$  in the
thermodynamic limit $V\rightarrow\infty$, when the sums over $\bf
{p}$ are transformed into the momentum integrals. Second, the
anomalous fluctuations of the particle number at the BEC point
correspond to the GCE description. In the CE and MCE, the number
of particles $N$ in a non-relativistic system is fixed by
definition, thus, $\omega_{c.e.}=\omega_{m.c.e.}=0$.
\\

\begin{figure}[t!]
 \hspace{-0.45cm}
 \includegraphics[width=0.5\textwidth]{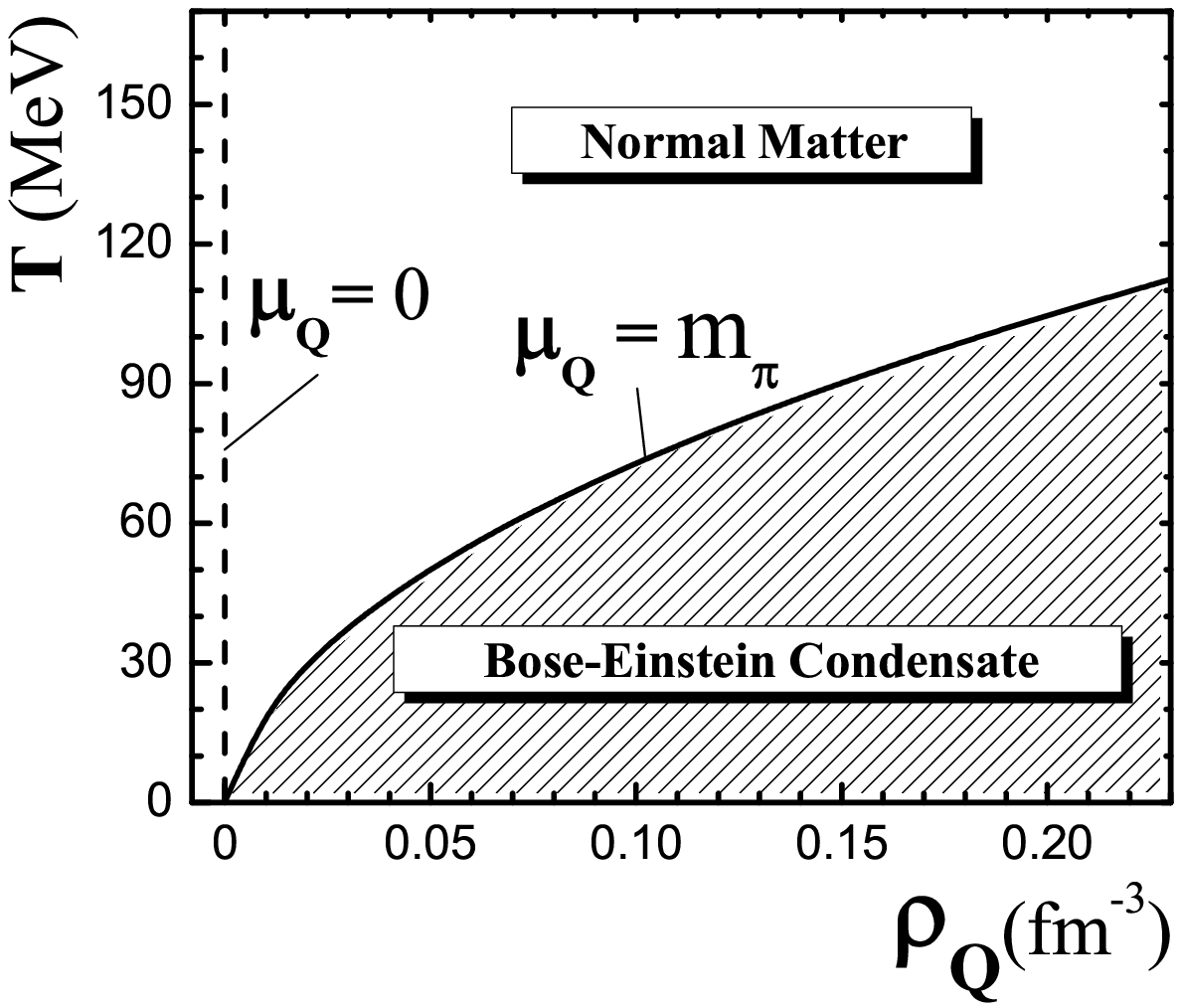}
 \includegraphics[width=0.52\textwidth]{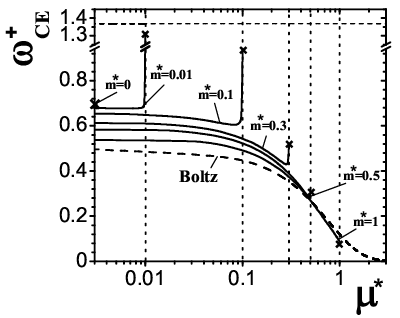}
 \vspace{-0.6cm}
 \caption{
{\it\bf Left:} The phase diagram of the relativistic ideal Bose
gas. The solid line  shows Bose condensation  temperature as a
function of the conserved charge density $\rho_Q$ given by Eq.~(6)
at $\mu_Q=m_{\pi}$. The dashed line shows $\rho_Q=\rho_+-\rho_-=0$
at $\mu_Q=0$.
 {\it\bf Right:}
The scaled variances $\omega_{c.e.}^{+}$ in pion Bose gas, are
shown as functions of $\mu^*\equiv\mu_Q/T$. The solid lines
present $\omega_{c.e.}^{+}$ at $m^*\equiv m_{\pi}/T =0.01,~ 0.1,~
0.3,~ 0.5,~ 1$.   The vertical dotted lines $\mu^*=m^*$
demonstrate the restriction $\mu^*\le m^*$ in the Bose gas. The
dashed horizontal line presents a value of
$\zeta(2)/\zeta(3)\simeq 1.368$ which is an upper limit for
$\omega_{c.e.}^{\pm}$ reached at $\mu^* =m^* \rightarrow 0$. The
crosses at $\mu^*=m^*$ correspond to the points of Bose
condensation. The crosses at $\mu^*=0$ correspond to
$\omega_{c.e.}^{\pm}(\mu^*=0,m^*\rightarrow 0$. The dashed line
corresponds to $\omega_{c.e.}^{+}$ in classical (Boltzmann) pion
gas (see Ref.~\cite{BF} for details).
 }\label{fig-RhoQ}
\end{figure}

The average values of the occupation numbers in the relativistic
ideal gas of pions equal to:
%
\eq{
\langle n_{{\bf p},j}\rangle~=~\frac{1}{\exp[(\sqrt{{\bf
p^2}+m_{\pi}^2}~-~\mu_j)/T]~-~1}~,
\label{np-pions}
}
where index $j$ enumerates 3 isospin pion states, $\pi^+,\pi^-$,
and $\pi^0$, the energy of one-particle states is taken as,
$\epsilon_{\bf p}=({\bf p^2}+m_{\pi}^2)^{1/2}$ with $m_{\pi}\cong
140$ MeV being the pion mass (we neglect a small difference
between the masses of charged and neutral pions). The inequality
$\mu_j\le m_{\pi}$ is a general restriction in the relativistic
Bose gas, and $\mu_{j}=m_{\pi}$ corresponds to the BEC. In
Ref.~\cite{BF} we discussed in details the Bose gas with one
conserved charge in the CE $(V,T,Q=const)$, i.e. the
$\pi^+\pi^-$-gas with fixed electric charge. This corresponds to
the GCE $(V,T,\mu_Q)$, thus, in Eq.~(\ref{np-pions}) $\mu_+
=\mu_Q$ and $\mu_-=-\mu_Q$ for $\pi^+$ and $\pi^-$, respectively.
Approaching the BEC of $\pi^+$ at $\mu_Q\rightarrow m_{\pi}$, one
finds the relation between $T_C$ and $\rho_Q\equiv\rho_+-\rho_-$
(see Fig.~\ref{fig-RhoQ}, Left). The picture of BEC of $\pi^-$ at
$Q<0$ and $\mu_Q\rightarrow -m_{\pi}$ is obtained by a mirror
reflection. BEC starts at $T=T_{C}$ when
$\mu_{Q}=\mu_Q^{max}=m_{\pi}\;$. It gives:
 \eq{\label{T_BCQ}
  \rho_{Q}(T=T_C,\mu_{Q}=m_{\pi})
 \;=\;
 \frac{T_{C}\,m_{\pi}^2}{\pi^2}
 \sum_{n=1}^{\infty}\frac{1}{n}\,
 K_2\left(n\,m_{\pi}/T_{C}\right)\,\sinh(n\,m_{\pi}/T_{C}) ,
  }
where $K_2$ is the modified Hankel function. At $T_C/m_{\pi}\ll
1$, Eq.~(\ref{T_BCQ}) gives, $T_C =
\frac{2\pi}{[\zeta(3/2)]^{2/3}}\, \rho_Q^{2/3}m_{\pi}^{-1}
 \cong  3.31 \rho_Q^{2/3}m_{\pi}^{-1}$, which coincides with the
non-relativistic formula (\ref{TC-nonrel}). In the
ultrarelativistic limit, $T_C/m_{\pi}\gg 1$, Eq.~(\ref{T_BCQ})
gives
\eq{\label{TC1}
 T_C~=~ \sqrt{3}~\rho_Q^{1/2}~m_{\pi}^{-1/2}~.
 }
The results presented in Eqs.~(\ref{T_BCQ}-\ref{TC1}) are well
known (see, e.g., Ref.~\cite{kapusta}). For this system, the
particle number fluctuations near the BEC line within GCE and CE
were recently studied  in Ref.~\cite{BF}.
The scaled variance $\omega^+\equiv\langle (\Delta
N_+)^2\rangle/\langle N_+\rangle$ in the GCE goes to infinity.
This is similar to the non-relativistic case. On the other hand,
the scaled variance for negative particles, $\omega^-\equiv\langle
(\Delta N_-)^2\rangle/\langle N_-\rangle$, remains finite and even
decreases with $\mu_Q$. The pion numbers $N_+$ and $N_-$ fluctuate
in the both GCE and CE. However, the exact conservation imposed in
the CE on the system charge, $Q=N_+-N_-$, suppresses anomalous
fluctuations at the BEC point: $\omega_{c.e.}^+$ (see
Fig.~\ref{fig-RhoQ}, Right) is finite with the upper limit,
$\zeta(2)/\zeta(3)\cong 1.368$ (see details in Ref.~\cite{BF}).
\\

A formation of the pion system with large electric charge density
$\rho_Q$ in high energy collisions does not look realistic.  In
what follows we discuss a rather different pion system which may
be created in high multiplicity events \cite{PBEC}. We use the MCE
$(V,E,Q=0,N_{\pi}=const)$ formulation, where the total system
energy $E$, electric charge $Q\equiv N_+ -N_-=0$, and total number
of pions,
 \eq{\label{Npi}
 N_{\pi} \;=\; N_0 \;+\; N_+ \;+\; N_-~,
 }
will be fixed. Such a system can be also described in  the
GCE~$(V,T,\mu_Q\!\!=\!0,\mu_{\pi})$ formulation, with
$\mu_+=\mu_{\pi}+\mu_Q$, $\mu_-=\mu_{\pi}-\mu_Q$, and
$\mu_0=\mu_{\pi}$ in Eq.~(\ref{np-pions}). We restrict
$\mu_Q\!\!=\!0$ and consider BEC when
$\mu_{\pi}\!\rightarrow\!m_{\pi}$. The $\mu_Q\!\!=\!0$ corresponds
to zero electric charge, $Q\!\!=\!0$ or $N_+=N_-$, in the pion
system.

\begin{figure}[h!]
 \hspace{-0.45cm}
 \includegraphics[width=0.5\textwidth]{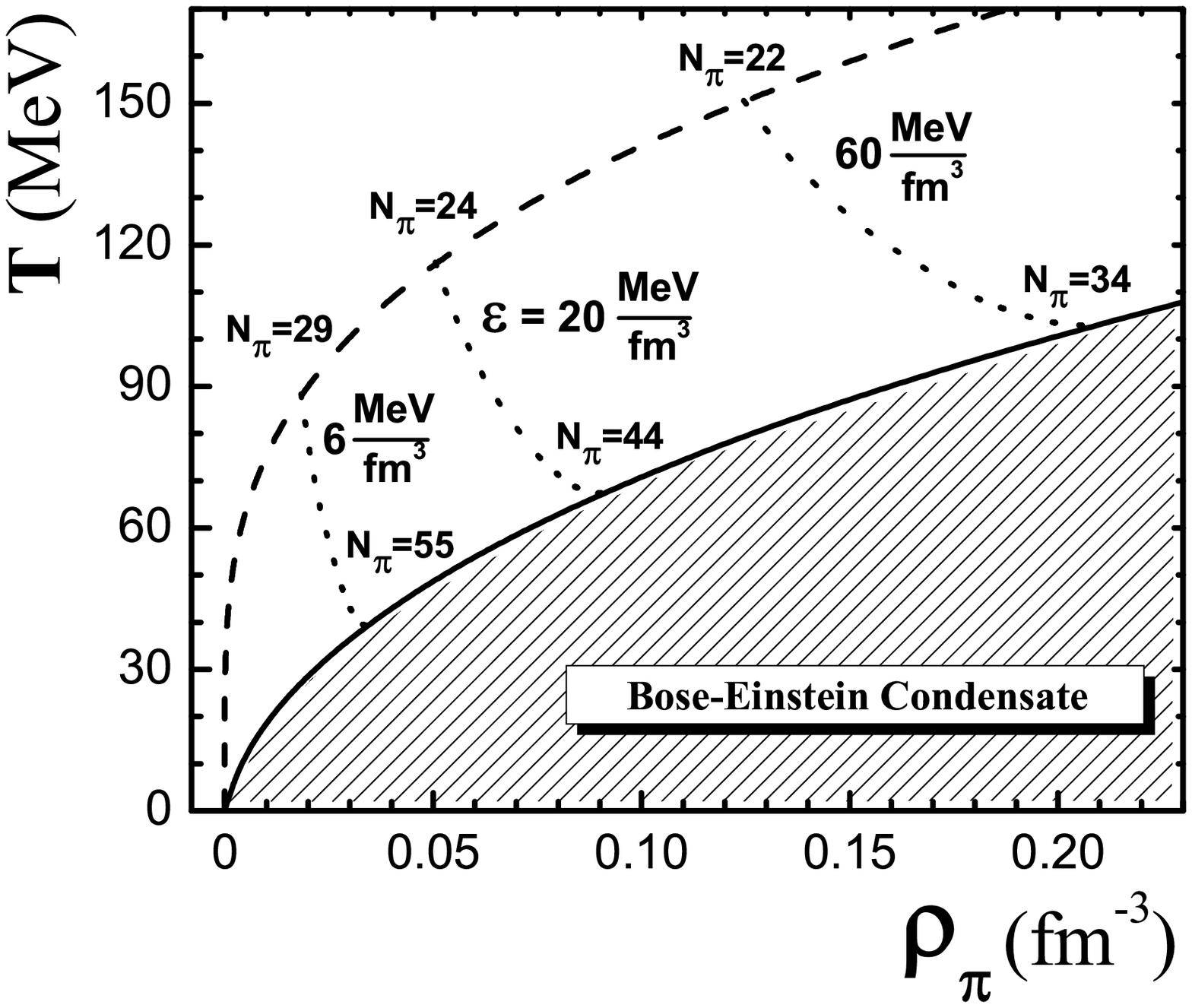}
 \includegraphics[width=0.5\textwidth]{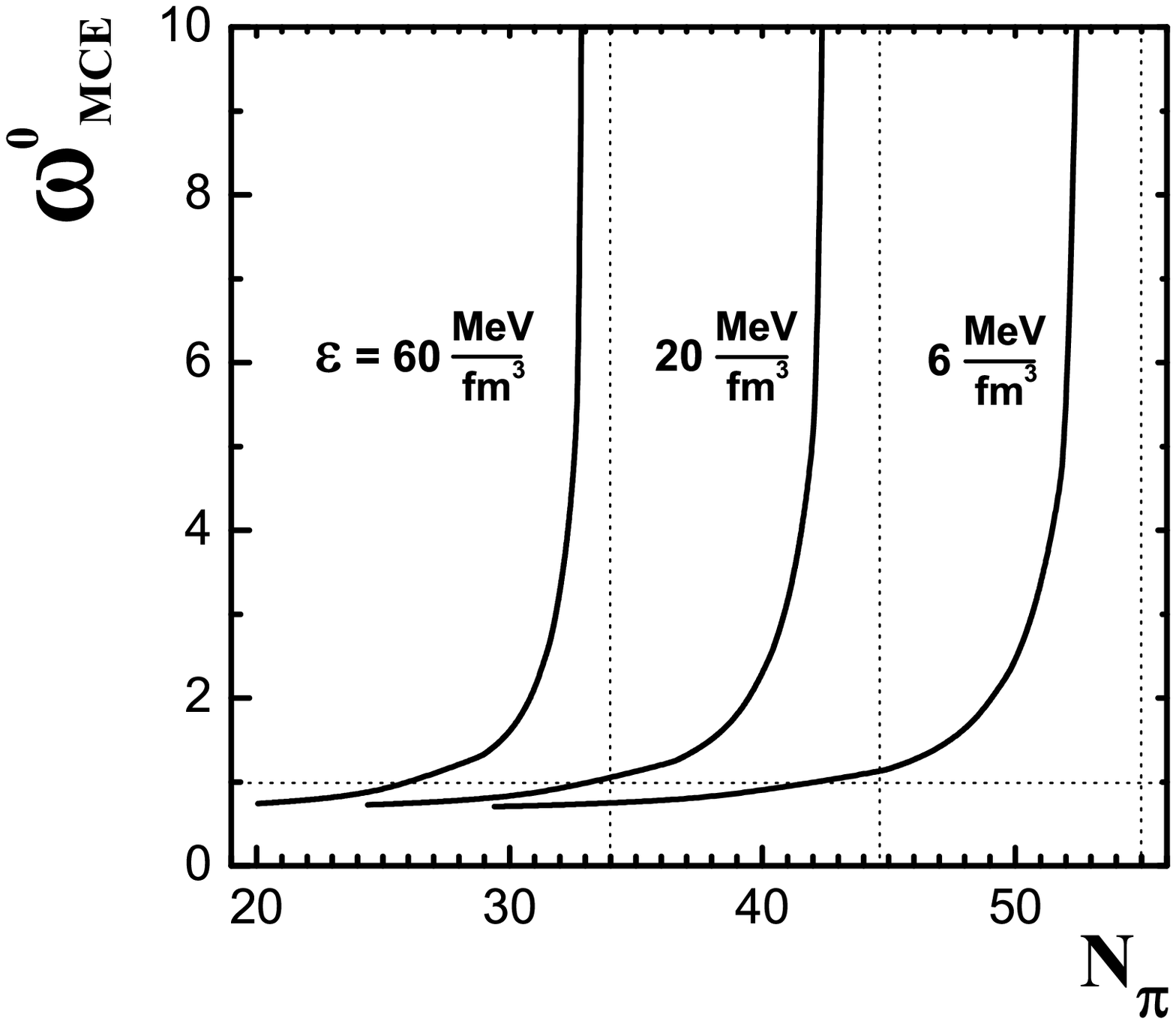}
 \vspace{-0.6cm}
 \caption{
{\it\bf Left:} The phase diagram of the pion gas with $\mu_Q=0$
\cite{PBEC}. The dashed line corresponds to
$\rho_{\pi}(T,\mu_{\pi}\!\!=0)$, and the solid line to BEC. The
dotted lines show the states
 with fixed energy densities: $\varepsilon = 6, 20, 60$ MeV/fm$^3$.
 The $N_{\pi}$  numbers in the figure correspond to $\mu_{\pi}=0$
 and $\mu_{\pi}=m_{\pi}$ at these energy densities for the total pion energy,
 $E~=~9.7$~GeV. 
 {\it\bf Right:} The scaled variance of neutral pions in the MCE  is
presented as the function of the total number of pions
\cite{PBEC}. Three solid lines correspond to different energy
densities: $\varepsilon = 6, 20, 60$ MeV/fm$^3$. The total energy
of the pion system is assumed to be fixed, $E= 9.7$~GeV. The
vertical dotted lines correspond to the points on the BEC line at
the specific values of the energy density. }\label{fig-cond}
\vspace{-0.5cm}
\end{figure}

The pion density is equal to
$\rho_{\pi}(T\!,\mu_{\pi})\!=\!\sum_{{\bf p},j} \langle n_{{\bf
p},j}\rangle/V$. The phase diagram of the ideal pion gas in
$\rho_{\pi}-T$ plane is presented in Fig.~\ref{fig-cond} (Left).
BEC starts at  $T=T_{C}$ when
$\mu_{\pi}=\mu^{max}_{\pi}=m_{\pi}\;$. It gives \cite{PBEC}:
 \eq{\label{T_BC}
  \rho_{\pi}(T=T_C,\mu_{\pi}=m_{\pi})
 \;=\;
  \frac{3\,T_{C}\,m_{\pi}^2}{2\pi^2}
 \sum_{n=1}^{\infty}\frac{1}{n}\,
 K_2\left(n\,m_{\pi}/T_{C}\right)\,\exp(n\,m_{\pi}/T_{C}) .
  }
 %
Note an essential difference between Eq.~(\ref{T_BC}) and
Eq.~(\ref{T_BCQ}): a presence of $\exp(nm_{\pi}/T_C)$ in
Eq.~(\ref{T_BC}), instead of $\sinh(nm_{\pi}/T_C)$ in
Eq.~(\ref{T_BCQ}). The Eq.~(\ref{T_BC}) gives the BEC line shown
by the solid line in Fig.~\ref{fig-cond} (Left). If
$T_{C}/m_{\pi}\ll 1$, from Eq.~(\ref{T_BC}) one finds,
$ T_{C} = 2\pi[3\zeta(3/2)]^{-2/3}\rho_{\pi}^{2/3}m_{\pi}^{-1}
 \cong 1.59\rho_{\pi}^{2/3}m_{\pi}^{-1}$.
%
This again corresponds to the non-relativistic limit
(\ref{TC-nonrel}) discussed above, but with a degeneracy factor
$g_{\pi}\!=\!3$. In the ultrarelativistic limit,
$T_{C}/m_{\pi}\!\gg\!1$, Eq.~(\ref{T_BC}) leads to a new relation
\cite{PBEC}:
 \eq{\label{TC2}
 T_{C}\!=\![\pi^2/3\zeta(3)]^{1/3}\rho_{\pi}^{1/3}~\cong
 1.4~\rho_{\pi}^{1/3}~,
 }
which differs from Eq.~(\ref{TC1}) and does not include the
dependence on $m_{\pi}$.

Let us consider the region in $\rho_{\pi}-T$ plane between the
$\mu_{\pi}\!=\!0$ and $\mu_{\pi}\!=\!m_{\pi}$ lines. The lines of
fixed energy density, $\varepsilon(T,\mu_{\pi})=\sum_{{\bf
p},j}\epsilon_{{\bf p}}\,\langle n_{{\bf p},j}\rangle/V$, are
shown as dotted lines in Fig.~\ref{fig-cond} (Left) inside this
region for three fixed values of $\varepsilon$. An increase of
$\rho_{\pi}$ at constant $\varepsilon$ leads to the increase of
$\mu_{\pi}$ and decrease of $T$. In this letter we discuss how the
system approaches the BEC line
$(\mu_{\pi}\!=\!m_{\pi},T\!=\!T_{C})$, and do not touch the region
$(\mu_{\pi}\!=\!m_{\pi},T\!<\!T_{C})$ below this line where the
non-zero BE condensate is formed. The GCE~$(V,T,\mu_Q,\mu_{\pi})$,
MCE~$(V,E,Q,N_{\pi})$, and CE~$(V,T,Q,N_{\pi})$ are equivalent for
average quantities, including average particle multiplicities, in
the thermodynamic limit. Thus, Eq.~(\ref{T_BC}) and phase diagram
in Fig.~1 remain the same in all statistical ensembles. However,
the pion number fluctuations are very different in different
ensembles. Before starting to calculate the pion number
fluctuations let us make several comments.

As an example we consider the high multiplicity events in $p+p$
collisions at IHEP (Protvino) accelerator with the beam energy of
70~GeV (see Ref.~\cite{Dubna} on the experimental project
``Thermalization'', team leader V.A.~Nikitin). In  the reaction
$p+p\rightarrow p+p+N_{\pi}$ with small final proton momenta in
the c.m.s.,  the total c.m. energy of created pions is $E\cong
\sqrt{s}-2m_{p}\cong 9.7$~GeV. The trigger system designed at JINR
(Dubna) selects the events with $N_{\pi}>20$ in this reaction.
This makes it possible to accumulate the samples of events with
fixed $N_{\pi}=30\div 50$ and the full pion identification during
the next 2 years \cite{nikitin}. Note that for this reaction the
kinematic limit is $N_{\pi}^{max}=E/m_{\pi}\cong 70$. We stress
that the IHEP experiment will measure the both charged and neutral
pions (see Ref.~\cite{pi-zero}). A reliable measurement of $N_0$
number in each event is a crucial point for the identification of
the BEC line suggested in this letter. The BEC signatures
discussed below become useless if the $\pi^0$ number cannot be
measured reliably.

The pion system in the thermal equilibrium is expected to be
formed for high multiplicities. The volume of the pion gas system
is estimated as, $V=E/\varepsilon(T,\mu_{\pi})$, and the number of
pions equals to $N_{\pi}= V\rho_{\pi}(T,\mu_{\pi})$. The values of
$N_{\pi}$ at $\mu_{\pi}=0$ and $\mu_{\pi}=m_{\pi}$ for 3 different
values of energy density $\varepsilon$ are shown in
Fig.~\ref{fig-cond} (Left) for the fixed total pion energy of
$E=9.7$~GeV. Note that the statistical approach to hadron
production in p+p collisions has been used successfully to
calculate the particle number ratios within the CE \cite{becCE}
and MCE \cite{becMCE}. Such an approach is usually applied to a
sample of the minimum bias events. Our suggestion has two new
points. First, it is a selection of the sample of events with high
pion multiplicity $N_{\pi}$. Most part of the available energy is
then spent to the pion production. Thus, a strong longitudinal
motion seen in the inclusive data is suppressed in high
multiplicity events because of the energy conservation. The pion
system may approach the state of global thermal equilibrium with
the thermodynamical parameters close to the BEC line. To search
the BEC effects we suggest to study the specific event-by-event
fluctuations of the number of pions. This is a second new point of
our suggestion.
%
%
%
%
%
%
%
The typical expected temperature of the pion gas approaching the
BEC line is about of $T= 60-90$~MeV (see Fig.~1). One can then
calculate the average thermal energy per particle, a rough
estimate gives: $3T/2 = 90-135$~MeV. On the other hand, a pion
from the $\rho$-meson decay has a much larger `kinetic energy' of
about 245~MeV in the $\rho$-meson rest frame. The pion energy
becomes even larger due to a presence of non-zero rho-meson
momenta in the c.m.s. of the p+p collision. Thus, we conclude that
for the high multiplicity events discussed in this letter a
presence of large number of resonances decaying into pions is
strongly suppressed because of the energy conservation.

For $Q=0$, the average pion multiplicities, $\langle N_0 \rangle
=\langle N_{\pm} \rangle =N_{\pi}/3$, are the same in all
statistical ensembles for large systems. This thermodynamic
equivalence is not, however, valid for the scaled variances of
pion fluctuations. The system with the fixed electric charge,
$Q=0$, the total pion number, $N_{\pi}$, and total energy of the
pion system, $E$, should be treated in the MCE. The volume $V$ is
one more (and unknown) MCE  parameter. The calculations below are
carried out in a large volume limit, thus, parameter $V$ does not
enter explicitly in the formulae for the scaled variances.

The microscopic correlators in the MCE ($V, E, Q=0, N_{\pi}$)
equal to (see also Refs.~\cite{PBEC,MCE,CLT}):
%
 \eq{ \label{mcorr}
 \langle \Delta  n_{{\bf p}}^j\, \Delta n_{{\bf k}}^j
\rangle_{m.c.e.}
  =  \upsilon_{{\bf p},j}^2~\delta_{{\bf p}{\bf k}}\delta_{ji}
  -  \frac{\upsilon_{{\bf p},j}^2\,\upsilon_{{\bf k},i}^2}
          {|A|}
   \Big[ q_jq_i\,M_{qq}
  +  M_{\pi\pi} + \epsilon_{{\bf p} i}\epsilon_{{\bf k}j}\, M_{\epsilon\epsilon}
 - (\epsilon_{{\bf p} i} + \epsilon_{{\bf k}j})\,M_{\pi\epsilon}
 \Big]\, ,
  }
where $q_+\!\!=\!1,\,q_-\!\!=\!-1,\,q_0\!\!=\!0$,
\eq{\label{vp2}
& \upsilon_{{\bf p},j}^2~=~ \upsilon_{\bf p}^2~=~
  \langle n_{\bf p}\rangle~  (1~+~\langle n_{\bf p}\rangle)~, \quad
& \langle n_{\bf{p}}\rangle~ = ~\{\exp[(\sqrt{{\bf
p^2}+m_{\pi}^2}-\mu_{\pi})/T]~-~1\}^{-1}~,
}
$|A|$ is the determinant and  $M_{ij}$ are the minors,
\eq{\label{minors}
 M_{qq}  = \Delta (\pi^2)\,\Delta (\epsilon^2)-(\Delta(\pi\epsilon))^2,
 &&
 M_{\pi\pi} = \Delta (q^2)\,\Delta (\epsilon^2),
 &&
 M_{\epsilon\epsilon} = \Delta (q^2)\,\Delta (\pi^2),
 &&
 M_{\pi\epsilon} = \Delta (q^2)\,\Delta (\pi\epsilon),
 }
of the correlation matrix $A$,
\eq{\label{A}
 A \;=\;
 \begin{pmatrix}
 \Delta (q^2) & 0 & 0\\
 0 & \Delta (\pi^2) & \Delta (\pi\epsilon)\\
 0 & \Delta (\pi\epsilon) & \Delta (\epsilon^2)
 \end{pmatrix}~.
 }
The matrix $A$ (\ref{A}) has the following elements:
 \eq{\label{elements}
  \Delta(q^2)
  &\;=\; \sum_{{\bf p},j}q_j^2 \upsilon_{{\bf p},j}^2
   \;=\; 2\sum_{\bf p}\upsilon_{\bf p}^2
   \;,
 &&
 \Delta (\pi^2)
   \;=\; \sum_{{\bf p},j} \upsilon_{{\bf p},j}^2
   \;=\; 3\sum_{\bf p}\upsilon_{\bf p}^2\;,
 \nonumber
 \\
 \Delta (\epsilon^2)
  &\;=\; \sum_{{\bf p},j} \epsilon_{\bf{p}}^2 \upsilon_{{\bf p},j}^2
   \;=\; 3\sum_{\bf p} \epsilon_{\bf{p}}^2 \upsilon_{\bf p}^2\;,
&&
 \Delta (\pi \epsilon)
  \;=\; \sum_{{\bf p},j} \epsilon_{\bf{p}} \upsilon_{{\bf p},j}^2
   \;=\; 3\sum_{\bf p} \epsilon_{\bf{p}} \upsilon_{\bf p}^2\;.
 }

Note that the first term in the r.h.s. of Eq.~(\ref{mcorr})
corresponds to the GCE. Correlations between differently charged
pions, $j\!\neq\! i$, and between different single modes, ${\bf
p}\!\ne\!{\bf k}$, are absent in the GCE:
%
\eq{
\omega^+ ~=~ \omega^-~ = \omega^0~
 \equiv \omega~ =~ 1~ + ~\frac{ \sum_{\bf{p}}\langle
n_{\bf{p}}\rangle^2}{\sum_{\bf{p}}\langle n_{\bf{p}}\rangle}~,
\label{omega}
}
similar to  non-relativistic result (\ref{omega-nonrel}), but with
$\langle n_{\bf{p}}\rangle$ given by the relativistic relation
(\ref{vp2}).
In the GCE the numbers $N_+$, $N_-$, and $N_0$ fluctuate
independently of each other. The Bose effects in the pion system
are small if $\mu_{\pi}=0$. For $\mu_{\pi}=0$, one finds
$\omega=1.01\div 1.12$ in the temperature interval $T=40 -
160$~MeV (note that $\omega=1$ in the Boltzmann approximation).
The Bose effects increase with $\mu_{\pi}$, and $\omega\rightarrow
\infty$ at $\mu_{\pi}\rightarrow m_{\pi}$, i.e. approaching the
BEC line the GCE calculations give anomalous fluctuations for
$N_+$, $N_-$, and $N_0$.

The MCE $(V,E,Q=0,N_{\pi}=const)$ formulation means the
restrictions of the exactly fixed total system energy $E$,
electric charge $Q=N_+ -N_-=0$, and total number of pions
$N_{\pi}$ (\ref{Npi}) for the each microscopic state of the
system. This changes the pion number fluctuations. From
Eq.~(\ref{mcorr}) one notices that the MCE fluctuations of each
mode ${\bf p}$ are reduced, and the (anti)correlations between
different modes ${\bf p}\ne {\bf k}$ and between different charge
states appear. This results in a suppression  of all scaled
variances $\omega^j_{m.c.e.}$ in comparison with the corresponding
ones $\omega$ in the GCE. A nice feature of the MCE microscopic
correlators (\ref{mcorr}) is that although being different from
that in the GCE, they are expressed with the quantities calculated
in the GCE. The MCE scaled variances depend on two GCE parameters:
$T$ and $\mu_{\pi}$.

The  substitution of Eqs.~(\ref{vp2}-\ref{elements}) in
Eq.~(\ref{mcorr}) and straightforward calculations lead to the
following MCE scaled variance for neutral pions \cite{PBEC}:
\eq{
 \omega^0_{m.c.e.}
  ~ = ~
 \frac{\sum_{\bf{p},\bf{k}}\langle
 \Delta n_{\bf{p}}^0~\Delta n_{\bf{k}}^0
 \rangle_{m.c.e.}}{\sum_{\bf{p}}\langle
 n_{\bf{p}}^0\rangle}~=~\frac{2}{3}~\omega~,
 %
 \label{omega-mce}
 }
where $\omega$ is given by Eq.~(\ref{omega}) and corresponds to
pion fluctuations in the GCE.
Due to the conditions, $N_+= N_-$ and $N_++N_-+N_0= N_{\pi}$, and
equal average multiplicities, $\langle N_0\rangle=\langle
N_+\rangle =\langle N_-\rangle =N_{\pi}/3$, it follows
\cite{PBEC}:
\eq{
\omega_{m.c.e.}^{\pm}~=~\frac{1}{4}~\omega_{m.c.e.}^0~=~\frac{1}{6}~\omega~,
&& \text{and} &&
\omega_{m.c.e.}^{ch}~=~\frac{1}{2}~\omega_{m.c.e.}^0~=~\frac{1}{3}~\omega~,
}
where $N_{ch}\equiv N_++N_-$. The behavior of $\omega^0_{m.c.e.}$
(\ref{omega-mce}) is shown in Fig.~\ref{fig-cond} (Right). To make
a correspondence with $N_{\pi}$ values, we consider again the $p+p
\rightarrow p+p+N_{\pi}$ collisions at the beam energy of 70~GeV
and take  the pion system energy to be equal to $E\!=\!9.7$ GeV.
Despite of the MCE suppression the scaled variances for the number
fluctuations of $\pi^0$ and $\pi^{\pm}$ increase dramatically and
abruptly when the system approaches the BEC line.

The following inequalities are always hold for particle number
fluctuations in different ensembles:
$\omega^j_{m.c.e.}<\omega^j_{c.e.}<\omega^j_{g.c.e.}$. Therefore,
if the anomalous BEC fluctuations are present in the MCE, they are
also exist (and even larger) in the CE and GCE. The reverse
statement is not true. The anomalous BEC fluctuations of the GCE
may disappear in the CE or MCE. We found that for the system with
$N_{\pi}=const$ and $Q=0$  the anomalous BEC fluctuations are not
washed out by exact conservation laws of the CE and MCE. This is
an advantage of the system with $N_{\pi}=const$ and $Q=0$. Let us
repeat again (see discussion just after Eq.~(\ref{TC1}) and
Ref.~\cite{BF}) that the anomalous BEC fluctuations at high charge
density $\rho_Q$ disappear in the CE and/or MCE. As another
instructive example let us consider the MCE
$(V,E,Q\!\!=\!0,N_{ch}=const)$, i.e. fixed $N_{ch}=N_++N_-$,
instead of fixed $N_{\pi}$ (\ref{Npi}). The corresponding GCE
formulation gives the following pion chemical potential:
$\mu_+=\mu_{\pi}$, $\mu_0=0$, $\mu_-= \mu_{\pi}$ in
Eq.~(\ref{np-pions}) ($\mu_Q=0$, as before, because of $Q=0$
condition). When $\mu_{\pi}\rightarrow m_{\pi}$ the system
approaches the BEC line for $\pi^+$ and $\pi^-$. The thermodynamic
behavior and position of this BEC line can be easily found.
Approaching the BEC line one can also find
$\omega^{\pm}\rightarrow\infty$ in the GCE. The pion number
fluctuations are, however, very different in the both CE
$(V,T,Q=0,N_{ch})$ and MCE $(V,E,Q=0,N_{ch})$. In the statistical
ensembles with fixed $N_{ch}$ and $Q$ no anomalous BEC
fluctuations are possible. The numbers of $N_+$ and $N_-$ are
completely fixed by the conditions $Q=N_+-N_-=0$ and
$N_{ch}=N_++N_-=const$, thus,
$\omega_{c.e.}^{\pm}=\omega_{m.c.e.}^{\pm}=0$. The number $N_0$
fluctuates, but $\mu_0=0$, thus, neutral pions are far away from
the BEC line and their fluctuations are small, $\omega^0\approx
1$, in all statistical ensemble formulations.

The broad distributions over $N_0$ and $N_{ch}$ close to the BEC
line also implies large fluctuations of the $f\equiv N_0/N_{ch}$
ratio. These large fluctuations were suggested (see, e.g.,
Ref.~\cite{DCC}) as a possible signal for the disoriented chiral
condensate (DCC). The DCC leads to the distribution of $f$ in the
form, $dW(f)/df=1/(2\sqrt{f})$. The thermal Bose gas corresponds
to the $f$-distribution centered at $f=1/2$. Therefore,
$f$-distributions from BEC and DCC are very different, and this
gives a possibility to distinguish between these two phenomena.

The calculations presented in this letter should be improved by
taking into account the finite size effects, pion-pion
interactions, and some other effects. Examples discussed in
Refs.~\cite{CE,MCE} demonstrate that the thermodynamical limit for
the average multiplicities and scaled variances is reached rather
quickly. Normally, if the average pion multiplicity is about of
$N_{\pi}=10$, the deviations from thermodynamic limit in the ideal
pion gas are only a few percents. The Bose-Einstein condensation
is a phase transition phenomenon. Thus, the infinite volume limit
is of a principal importance. A strict mathematical meaning of the
phase transition (and its order) has only sense in the infinite
volume limit. Of course, the real systems are finite. For our
applications this means that the scaled variance for neutral pions
shown in Fig. 2 does not increase up to `infinity', but it is
restricted from above in the finite system. A detailed study of
the finite size effects for the BEC is now under investigation and
will be published elsewhere. The BEC temperature of about $ T=
60-90$~MeV corresponds to the pion number density  of
$\rho_{\pi}=0.1-0.15$~fm$^{-3}$ (see Fig. 1). This particle
density is not too large (it is smaller than the normal nuclear
density). This probably may justify the ideal pion gas
approximation considered in the present letter as a first step in
the modelling of multi-pion states. The effects of pion
interactions will be discussed in the future studies. Preliminary
estimates suggest that the BEC signatures suggested in the present
letter may survive the complications. A crucial point is the
analysis of the samples of high $N_{\pi}$ events. The required
$N_{\pi}$ values for the BEC are much larger than the average pion
multiplicity per collision, thus, these high $N_{\pi}$ events are
rather rare and give negligible contributions to inclusive
observables in high energy collisions.
With increasing of $N_{\pi}$ in the sample with fixed total
energy, the temperature of the pion system has to decrease and it
approaches the BEC line. This can happen in different ways: at
constant energy density $\varepsilon$, at constant pion density
$\rho_{\pi}$, or with decreasing of both $\varepsilon$ and
$\rho_{\pi}$. The pion system should move to the BEC line one way
or another. In the vicinity of the BEC line (no BE condensate is
yet formed) one observes an abrupt and anomalous increase of the
scaled variances of neutral and charged pion number fluctuations.
This could (may be even should) be checked experimentally.

{\bf Acknowledgments.} We would like to thank F. Becattini,
K.A.~Bugaev, A.I.~Bugrij, I.M.~Dre\-min, M.~Ga\'zdzicki,
W.~Greiner, K.A.~Gridnev, M.~Hauer, I.N.~Mishustin,
St.~Mr\'owczy\'nski and Yu.M.~Si\-nyu\-kov for discussions and
comments. We are also grateful to E.S.~Kokoulina and
V.A.~Ni\-ki\-tin. They informed us on the experimental project
\cite{Dubna}, and this stimulated the present study. We thank
S.V.~Chubakova for help in the preparation of the manuscript.
V.~Begun would like to thank for the support the organizers of the
conference and The International Association for the Promotion of
Cooperation with Scientists from the New Independent states of the
Former Soviet Union (INTAS), Ref. Nr. 06-1000014-6454.

\end{document}